\documentclass[12pt]{article}

\usepackage{amsmath}
\usepackage{amssymb}
\usepackage{graphicx}
\usepackage{url}
\usepackage{multirow}
\usepackage{authblk}
\usepackage{mathtools}
\usepackage{hyperref}
\usepackage[margin=1in]{geometry}
\usepackage{natbib}
\usepackage{subfig}
\usepackage{graphicx}
\usepackage{tikz}
\usepackage{pgfplots}
\usepackage{blkarray}
\usepackage{adjustbox}

\pgfplotsset{compat=1.3}
\usepgfplotslibrary{groupplots}
\usetikzlibrary{calc}
\usetikzlibrary{positioning}
\usetikzlibrary{arrows}
\usetikzlibrary{shapes,snakes}
\DeclarePairedDelimiter\ceil{\lceil}{\rceil}
\DeclarePairedDelimiter\floor{\lfloor}{\rfloor}

\begin{document}

\title{Estimation of marriage incidence rates by combining two cross-sectional retrospective designs: Event history analysis of two dependent processes}

\author[1]{Sangita Kulathinal\footnote{Correspondence to: Sangita Kulathinal, Department of Mathematics and Statistics, University of Helsinki, PL 68 (Pietari Kalmin katu 5) 00014 Finland,  e-mail: \url{sangita.kulathinal@helsinki.fi}, Tel.: +358505608064.}}
\author[2]{Minna S\"a\"av\"al\"a}
\author[3]{Kari Auranen}
\author[4]{Olli Saarela}

\affil[1]{Department of Mathematics and Statistics, University of Helsinki, Finland}
\affil[2]{Population Research Institute, V\"aest\"oliitto, Helsinki, Finland}
\affil[3]{Department of Mathematics and Statistics, University of Turku, Finland}
\affil[4]{Dalla Lana School of Public Health, University of Toronto, Canada}

\maketitle

\begin{abstract}
The aim of this work is to develop methods for studying the determinants of marriage incidence using marriage histories collected under two different types of retrospective cross-sectional study designs. These designs are: sampling of ever married women before the cross-section, a prevalent cohort, and sampling of women irrespective of marital status, a general cross-sectional cohort. While retrospective histories from a prevalent cohort do not identify incidence rates without parametric modelling assumptions, the rates can be identified when combined with data from a general cohort. Moreover, education, a strong endogenous covariate, and marriage processes are correlated. Hence, they need to be modelled jointly in order to estimate the marriage incidence. For this purpose, we specify a multi-state model and propose a likelihood-based estimation method. We outline the assumptions under which a likelihood expression involving only marriage incidence parameters can be derived. This is of particular interest when either retrospective education histories are not available or related parameters are not of  interest. Our simulation results confirm the gain in efficiency by combining data from the two designs, while demonstrating how the parameter estimates are affected by violations of the assumptions used in deriving the simplified likelihood expressions. Two Indian National Family Health Surveys are used as motivation for the methodological development and to demonstrate the application of the methods.
\end{abstract}

\section{Introduction}\label{section:intro}

In sociology and demography, population-based cross-sectional surveys have been used to estimate rates of events such as marriage or cohabitation. For estimation of marriage incidence rates, retrospective marriage history, e.g. age at first marriage, can be  collected by sampling at a cross section. Two commonly employed sampling designs at a cross-section are; (i) sampling of ever married women before the cross-section, a prevalent cohort, and (ii) sampling of women irrespective of marital status, a general cross-sectional cohort. Marriage histories are collected retrospectively under the two designs. We refer to studies based on these two designs as retrospective cohort studies I and II, respectively.

Similar designs are used in epidemiology to estimate incidence rate of a disease based on retrospective disease histories, with methods described in e.g. \citet{keiding:1991,keiding:2012}. \cite{keiding:2006} gives an overview of event history analysis and the cross-section with focus on complex sampling patterns. Further, \citet{saarela:2009} proposed combining retrospective event histories from individuals with prevalent disease and prospective follow-up of disease free individuals at the cross-section, incident cohort, to improve efficiency in estimating effects of time-invariant covariates on disease incidence. Gain in efficiency has also been demonstrated in estimation of survival time from disease onset to death based on combined prevalent and incident cohort data \citep{ning:2017, wolfson:2019}.

Although incidence rates estimation methods using retrospective disease histories are known in epidemiology, their application in other fields are sparse. In the sociological context, retrospective event histories are typically collected under the cross-sectional retrospective designs described earlier. To estimate incidence of the outcome when the outcome of interest is correlated with an endogenous covariate process, the outcome and the covariate processes need to be modelled jointly. Moreover, the estimation method should account for the sampling. In the absence of complete covariate process histories at the cross-section, incidence rates estimation may be possible only under special assumptions or sufficient background information on the covariate processes.

The novelty of the present work is in modelling marriage  and education processes jointly using a multi-state model by combining the two retrospective cohort studies. We thus extend the existing likelihood-based methods for estimation of incidence rates to simultaneously account for two different sampling patterns; two correlated processes; and two time scales. We outline the assumptions under which the likelihood expressions for the marriage incidence rates can be derived when complete retrospective histories of the education process are not available or when parameters characterizing the education process are not of  interest. In a simulation study, we assess the gain in efficiency due to using the proposed method over relying on data from either of the two studies. We apply the methods to two nationally representative Indian National Family Health Surveys (NFHS) data to study the trends and determinants of marriage incidence in India.  While we present results in the context of education and marriage, the results are general and can be applied to other similar settings.

The paper is organised as follows. Section \ref{section:data} introduces the empirical data from the two NFHS. Section \ref{section:women} outlines the model of female marriage incidence and derives the necessary likelihood expression of the model parameters to estimate them from cross-sectional data. Section \ref{section:predictive} considers calculation of predictive probabilities based on the model. A simulation study and data analysis results are presented in sections \ref{section:simulation} and \ref{section:results}. The paper concludes with a discussion. 

\section{The data}\label{section:data} 
\begin{figure}[!h]
\centerline{\includegraphics[height=8cm,width=8cm]{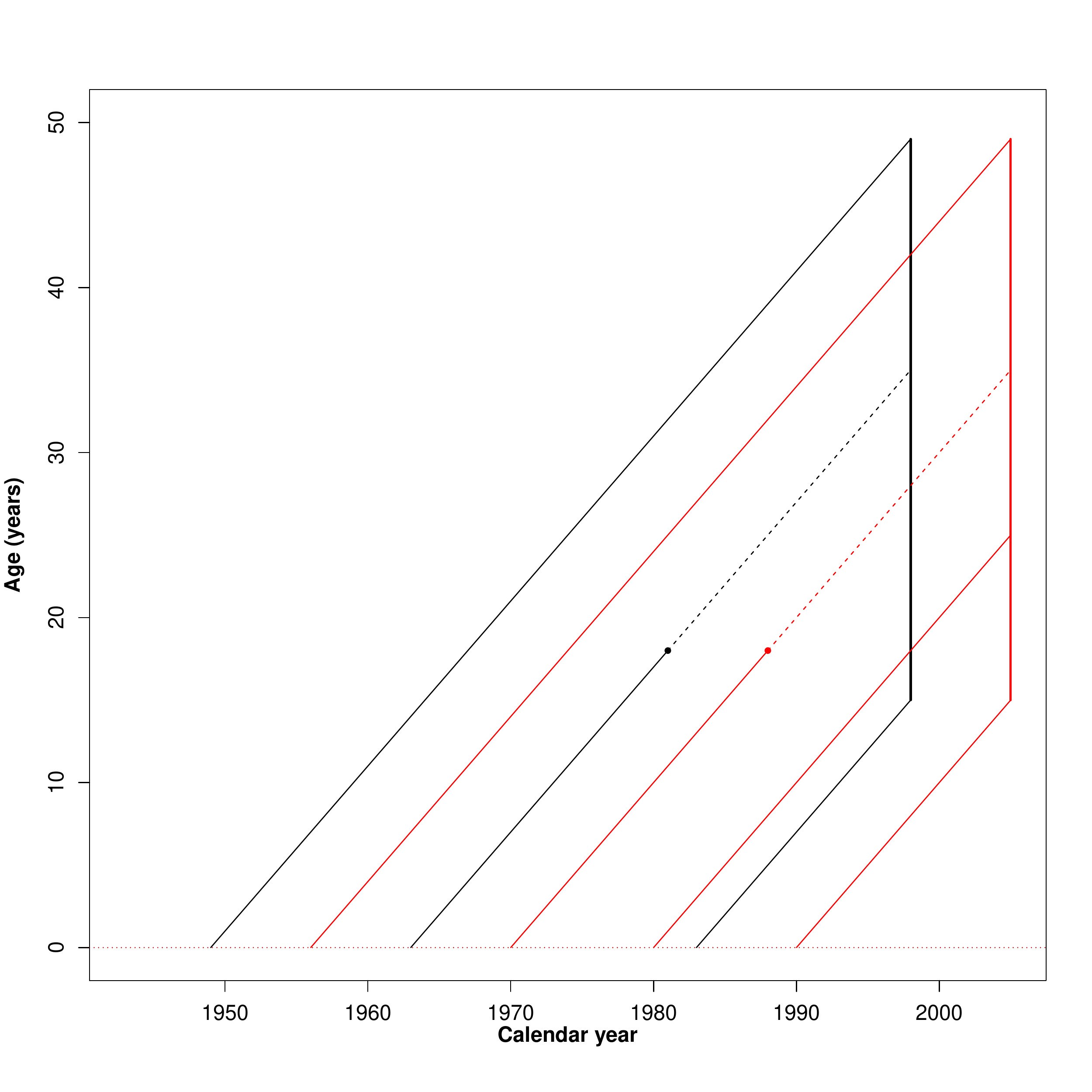}}
\caption{Lexis diagram illustration of the two cross-sectional surveys. NFHS-2 (black) collected retrospective marriage histories  
from ever married women only (retrospective cohort study I). NFHS-3 (red) included also never married women  and collected retrospective marriage histories from currently married women (retrospective cohort study II).}\label{figure:lexis}
\end{figure}

The motivation for this work comes from the estimation of marriage incidence rates and their determinants using two NFHS; surveys conducted in India during 1998-99 (NFHS-2) and 2005-06 (NFHS-3). The NFHS-2, an example of retrospective cohort study I,  was a cross section of a nationally representative sample of 91196 households with 90265 ever-married women aged 15-49 years and gave a retrospective cohort of ever married women. The NFHS-3, an example of retrospective cohort study II, included 109041 households with 124373 women aged 15-49 years irrespective of marital status and gave a retrospective cohort, irrespective of the current status of marriage at the time of survey. The data and reports of the NFHS are available through the National Family Health Survey website (\url{http://rchiips.org/nfhs/}). A schematic Lexis diagram illustration of the three cohorts is presented in Figure \ref{figure:lexis}. Education is known to be a key determinant of marriage and hence, we model the joint dependency of the education and marriage processes in this context. The NFHS reports clearly bring out differences between the states
with respect to education  (cf. Appendix A). Hence, we concentrate on, in addition to whole India, a subset of the NFHS data that covers four Indian states (Kerala, Maharashtra, Punjab and Rajasthan) with different demographic situations.

The data used in the current analysis include each female participant's age at the time of the survey, age at first marriage, state, birth cohort, urban/rural residence, caste, religion, and highest educational level completed, categorized as in Table~\ref{table:covariates}. The total number of study subjects in the four states was 38052 (Table~\ref{table:descriptivewomen}, Appendix A).

\begin{table}[!h]
\small\sf\centering
\caption{Covariates in the marriage incidence model. The reference categories
are indicated as 'ref.'}
\centering
\begin{tabular}{llc}
\hline
Covariate & Category & Notation  \\
\hline
Birth cohort & 1942-62 (ref.) & $x_1=0$   \\
& 1962-72 &  $x_1=1$ \\
& 1972-82 &  $x_1=2$\\
& 1982-92 & $x_1=3$\\ \\
Residence status & Urban (ref.) & $x_2=0$  \\
& Rural &  $x_2=1$ \\ \\
Caste & Scheduled Caste (SC, ref.) & $x_3=0$  \\
& Scheduled Tribe (ST) & $x_3=1$ \\
& Other Backward Class (OBC) & $x_3=2$ \\
& Other & $x_3=3$ \\ \\
Religion & Hindu (ref.) & $x_4=0$ \\
& Muslim &  $x_4=1$ \\
& Christian & $x_4=2$ \\ 
& Sikh & $x_4=3$ \\ 
& Other & $x_4=4$ \\ \\
Education$^a$ &  None ($< 5$ years) (ref.) & $x_5=0$  \\
& Primary (5-9 years) &  $x_5=1$ \\
& Secondary (10-12 years) & $x_5=2$ \\ 
& Higher ($> 12$ years) & $x_5=3$ \\ \\
\hline
\end{tabular}
\begin{flushleft}Note: $^a$ ordinal variable\end{flushleft}
\label{table:covariates}
\end{table}

\vspace{.1in} 
\section{Joint modelling of education and marriage processes and estimation of marriage incidence rates}\label{section:women}
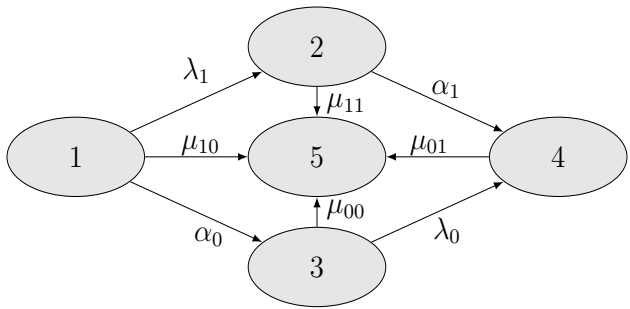
\begin{figure}[!h]
\begin{center}
\resizebox{0.5\textwidth}{4cm}{%
\begin{tikzpicture}[
    sharp corners=2pt,
    inner sep=7pt,
    node distance=5cm,
    >=latex]
\tikzstyle{my node}=[draw,minimum height=1cm,minimum width=2cm,fill=gray!20]
\node[ellipse,my node] (1){1};
\node[ellipse,my node,right=of 1](4){4};
\node[ellipse,my node] at ($(1)!0.5!(4)-(0pt,-1.5cm)$) (2) {2};
\node[ellipse,my node] at ($(1)!0.5!(4)-(0pt,0cm)$) (5) {5};
\node[ellipse,my node] at ($(1)!0.5!(4)-(0pt,1.5cm)$) (3) {3};
\draw[->] (1) -- (2);
\draw[->] (1) -- (5);
\draw[->] (1) -- (3);
\draw[->] (2) -- (5);
\draw[->] (2) -- (4);
\draw[->] (3) -- (5);
\draw[->] (3) -- (4);
\draw[->] (4) -- (5);
\path [-latex'] (1) -- node [text width=2.5cm,midway,above,align=center ] {$\lambda_1$} (2);
\path [-latex'] (1) -- node [text width=2.5cm,midway,below,align=center,xshift=1.0ex,yshift=0.2ex ] {$\alpha_0$} (3);
\path [-latex'] (1) -- node [text width=2.5cm,midway,below,align=right,xshift=-5ex,yshift=3ex ] {$\mu_{10}$} (5);
\path [-latex'] (2) -- node [text width=2.5cm,midway,below,align=right,xshift=-5ex,yshift=3ex ] {$\alpha_1$} (4);
\path [-latex'] (2) -- node [text width=2.5cm,midway,below,align=right,xshift=-3ex,yshift=2.0ex ] {$\mu_{11}$} (5);
\path [-latex'] (3) -- node [text width=2.5cm,midway,below,align=right,xshift=-5ex,yshift=1ex ] {$\lambda_0$} (4);
\path [-latex'] (3) -- node [text width=2.5cm,midway,below,align=right,xshift=-3ex,yshift=2.3ex ] {$\mu_{00}$} (5);
\path [-latex'] (4) -- node [text width=2.5cm,midway,below,align=right,xshift=-6ex,yshift=3.0ex ] {$\mu_{01}$} (5);
%\draw[<-] (healthy.355) -- (sick.185);
\end{tikzpicture}%
}
\caption{At-school and marriage processes as a multi-state model (states are: 1 = at school and unmarried, 2 = at school and married, 3 = out of school and unmarried, 4 = out of school and married, 5 = dead)}\label{figure:multistate}
\end{center}
\end{figure}

As noted earlier, education is known to be a key determinant of marriage and vice versa, and hence, are highly correlated with each other.  We model the two correlated processes; at-school and marriage processes,  in a multi-state modelling framework. Each process has two states indicating respective status, and the joint process can be described using a multi-state model as depicted in Figure~\ref{figure:multistate}. The state space of the joint process is  \{\emph{at school and unmarried, at school and married, out of school and unmarried, out of school and married, dead} \}. We denote these five states as $\{1, 2, \ldots, 5\}$, respectively. Of note, the at-school process jumps to \emph{out of school} state when the formal education ends. Let $a_e$ and $a_0$ be the minimum age of starting basic compulsory education and the minimum marriageable age $a_0 (> a_e)$, respectively. In the Indian context, $(a_e, a_0)$ are taken as $(6, 12).$

\pgfkeys{
   /pgf/number format/.cd, 
      set decimal separator={,{\!}},
      set thousands separator={}
}
\pgfplotsset{
   every axis/.append style = {
      line width = 1pt,
      tick style = {line width=1pt},
      label style={font=\small},
      tick label style={font=\footnotesize}
   }
}

\begin{figure}[!h]
\begin{center}
\resizebox{0.6\textwidth}{6cm}{%
\begin{tikzpicture}
    % provide shared options here with pgfplotsset:
    \pgfplotsset{
        height=6cm, width=9cm,
        no markers
    }
    % this is the leftmost y axis (y2)
    \begin{axis}[
        xmin=0,xmax=40,%--- CF
        xshift=-2.0cm,%-- CF
        width=2cm,
        hide x axis,
        axis y line*=left,
        ymin=0, ymax=21,
        ytick = {0,4,7,10,12,15,17,21},
        ylabel={\color{red} Schooling years ($\int_{0 \le u\le a} N_1(t(u),u) \,\textrm du$)}
    ]
    \end{axis}
    % this is the inner y axis (schooling status)
    \begin{axis}[
        xmin=0, xmax=10,
        xshift=-0.3cm,%-- CF
        width=2cm,
        hide x axis,
        axis y line*=left,
        ymin=0, ymax=1,
        ytick = {0,1},
        ylabel={\color{blue} Education status ($N_1(t,a)$)}
    ]
    \end{axis}
    % this is the unique x-axis
    \begin{axis}[
        height=2cm, yshift=-0.4cm,
        xmin=0, xmax=40,
        ymin=0, ymax=21,
        xtick = {0,6,10,13,18,21,25,30,35,40},
        ytick = {0,4,7,10,12,15,17,21},
        axis x line*=bottom,
        hide y axis,
        xlabel={Age ($a$ in years)}
    ]
    \end{axis}
    % this is the red curve
    \begin{axis}[
        xmin=0, xmax=40,
        ymin=0, ymax=21,
        hide x axis,
        hide y axis, 
    ]
        \addplot[dashed, red, domain=0:10] coordinates { (0,0) (6,0) };
        \addplot[very thick, red, domain=0:10] coordinates { (6,0) (18,12) };
        \addplot[very thick, red, domain=0:10] coordinates { (18,12) (40,12) };
%        \draw (axis cs:2,3) -- node[left]{Text} (axis cs:2,6);
    \end{axis}
    % this is the green curve
    \begin{axis}[
        xmin=0, xmax=40,
        ymin=0, ymax=1,
        hide x axis,
        hide y axis, 
    ]
        \addplot[dashed, green!75!black, domain=0:10] coordinates { (0,0) (12,0) };    
        \addplot[very thick, green!75!black, domain=0:10] coordinates { (12,0) (21,0) };
        \addplot[very thick, green!75!black, domain=0:10]  coordinates { (21,1) (40,1) };
    \end{axis}
    % this is the blue curve
    \begin{axis}[
        xmin=0, xmax=40,
        ymin=0, ymax=1,
        hide x axis,
        hide y axis, 
    ]
        \addplot[dashed, blue, domain=0:10] coordinates { (0,0) (6,0) };    
        \addplot[very thick, blue, domain=0:10] coordinates { (6,1) (18,1) };
        \addplot[very thick, blue, domain=0:10]  coordinates { (18,0) (40,0) };
    \end{axis}    
        % this is the black vertical line at the time of the marriage
    \begin{axis}[
        xmin=0, xmax=40,
        ymin=0, ymax=1,
        hide x axis,
        hide y axis, 
    ]
        \addplot[dotted, black, domain=0:10] coordinates { (12,0) (12,1) };
        \addplot[dotted, black, domain=0:10] coordinates { (21,0) (21,1) };
%        \draw (axis cs:2,3) -- node[left]{Text} (axis cs:2,6);
    \end{axis}

        % this is the black solid vertical line at the time of the cross-section
    \begin{axis}[
        xmin=0, xmax=40,
        ymin=0, ymax=1,
        hide x axis,
        hide y axis, 
    ]
        \addplot[solid, black, domain=0:10] coordinates { (25,0) (25,1) };
%        \draw (axis cs:2,3) -- node[left]{Text} (axis cs:2,6);
    \end{axis}

    % this is the right-hand y-axis (y3)
    \pgfplotsset{every axis y label/.append style={rotate=180}}
    \begin{axis}[
        xmin=0, xmax=10,
        ymin=0, ymax=1,
        xshift=0.3cm,%-- CF
        hide x axis,
        axis y line*=right,
        ytick = {0,1},
        ylabel={\color{green!75!black} Marital status ($N_2(t,a)$)}
    ]
    \end{axis}
\end{tikzpicture}%
}
\caption{A sample path of at-school (blue) and marriage (green) processes. Basic education starts at the age of 6 years and marriageable age is 12 years. The inner left y-axis indicates education status 0 (= out of school) and 1 (= at school) and the outer left y-axis gives the accumulated schooling years. The right y-axis is the marital status axis, 0 (= unmarried) and 1 (= married). Dashed black lines indicate observation period relevant for marriage process and solid black line indicates the cross-section.}\label{figure:samplepath}
\end{center}
\end{figure}
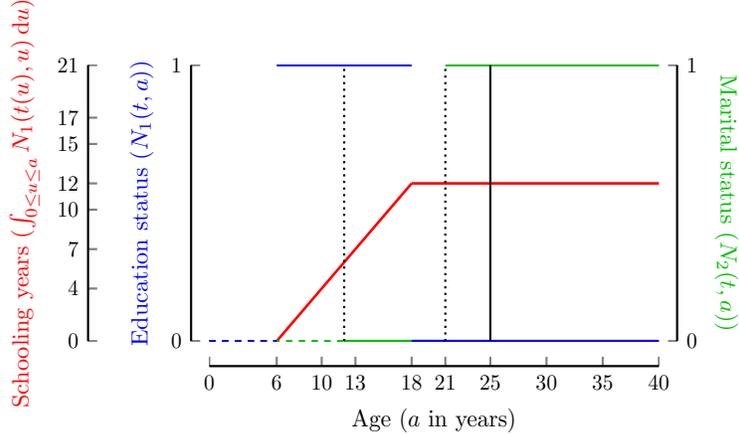

We denote the calendar time corresponding to age $a$ as $t(a) = t_0 + a$ where $t_0$ is the birth year and define both processes in a Lexis diagram with calendar time and age as the two time scales. We define the at-school process $\{N_1(t, a), a \ge 0, t = t(a)\}$ as a stochastic process giving the education status with $N_1(t, a)=1$ indicating being in school and $N_1(t, a)=0$ having stopped formal education (out of school) by age $a$ at time $t(a)$. Similarly, the marriage process $\{N_2(t, a), a \ge 0, t = t(a)\}$ is a stochastic process giving the marital status of a woman aged $a$ at time $t(a)$, with $N_2(t, a)=0$ indicating unmarried and $N_2(t, a)=1$ married status. The corresponding histories are defined as $\mathcal{F}_r(t, a) = \{N_r(s, u), u \le a, s(u) \le t(a)\}$, $r = 1, 2,$  respectively and the joint history as $\mathcal{F}(t,a) = \{(N_1(s,u), N_2(s,u)), u\le a, s\le t\}.$. 

The counting process $N_1(t, a)$ remains at zero between the age $0$ and $a_e$, that is between the birth year $t_0$ and the year $t(a_e)$. %and it remains at the maximum schooling year from the age at that attainment (and corresponding calendar period) till the age at death (death period).
Because of minimum marriageable age $a_0 (> a_e)$, the process $N_2(t,a)$ is zero for all $a < a_0$ and $t < t(a_0).$ The association between the two processes is modelled through the dependence on the joint history $\mathcal{F}(t,a).$ Because the two times  grow together with the same pace we denote the history using only one time scale as $\mathcal{F}(a)$. Time invariant information or fixed covariates at birth $(x)$ such as religion and caste are also included in this history. We also construct a deterministic counting process giving schooling years of a woman aged $a$ at time $t(a)$, as the accumulated history of at-school process $\{\int_{0 \le u\le a} N_1(t(u),u) \,\textrm du\}$.

The corresponding transition intensities for the two processes are defined as
\begin{align}
\MoveEqLeft \alpha_{1,k}(t,a\mid \mathcal{F}(a^-)) \nonumber \\ 
& =   \lim_{\Delta t\to 0}\frac{P(N_{1}(t+\Delta t, a+\Delta t) = 0 \mid  N(t^-, a^-)= (1,k), \mathcal{F}(a^-))}{\Delta t}, \nonumber
\end{align}  
\begin{align}  
\MoveEqLeft \lambda_{k,0}(t,a\mid \mathcal{F}(a^-))   \nonumber \\
& = \lim_{\Delta t\to 0}\frac{P(N_{2}(t+\Delta t, a+\Delta t) = 1 \mid N(t^-, a^-) = (k,0),  \mathcal{F}(a^-))}{\Delta t}, \nonumber
\end{align}
where $N(t,a) = (N_1(t,a), N_2(t,a))$ and $k = 0, 1$. Since the process $N_1(t, a)=1$ until the transition happens, we drop the subscript $1$ and simplify the notation  $\alpha_{1,k}(t,a\mid \mathcal{F}(a^-))$ to  $\alpha_{k}(t, a)$. 
Similarly, the process $N_2(t, a)=0$ until the transition happens, and hence, we use $\lambda_{k}(t, a\mid \mathcal{F}(a^-))$  as a simplified notation for $\lambda_{k,0}(t,a)$. Furthermore, we define $\mu_{jk}(t,a)$ to be the rate of moving to state 5 (dead), where $j, k \in \{0,1\}$ represent the current schooling and marriage status, respectively.

Figure~\ref{figure:samplepath} exhibits an example sample paths of the two processes based on the retrospective information collected at the   cross-sectional age of $25$. In the example, the formal school ends at the age of $18$ and marriage takes place at the age of $21$.   The at-school process remains in state 1 between the age of 6 years ($a_e=6$) and 18 years, then jumps to state 0. The marriage process starts at the age of 12 years ($a_0=12$) in state unmarried (0) and jumps to state married (1) at the age of 21 years. In addition, the deterministic counting process giving the accumulated schooling years is also shown. The observation process stops at the first marriage and our main interest in the joint processes is in the time interval between the age $a_0$ and the age at the first marriage. In the observation period, the multi-state process starts in state 1 at age 12 and calendar year $(t_0+12)$ and moves to state 3 before transitioning to state 4 at age $21$ and calendar year $(t_0+21)$. The schooling years at that time are 12 which are attained at the age of 18. In principle, changes in the at-school 
process after marriage can be inferred based on the method given below subject to the availability of data.

We derive the likelihood contributions for all possible event histories conditional on being alive in either state 1 or 3 at age $a_e$. Let us first consider an individual born in the calendar time $t_0$ and in state 1 (at school and unmarried) at age $a_e$. Figure~\ref{figure:multistate} shows possible transitions between the five states. We develop the model following notation of \citet{keiding:1991}, extending it to include two correlated processes. The probability density of being unmarried and at school with $w = z - a_e$ years of schooling and aged $[z, z+\textrm dz)$ at time $t$, is proportional to $\beta_1(t-z, a_e) k_1(t, z, w)$, where $\beta_1(t-z,a_e)$ is the probability density of being born in year $t-z$ and being in state 1 at age $a_e$ and
\begin{align}
k_1(t,z,w)  &=  \exp\left\{-\int_{a_e}^{z} [\mu_{10}(t - z + u, u) + \alpha_0(t - z + u, u)] \,\textrm du \right\}  \nonumber \\ 
&\quad\times \exp\left\{-\int_{a_0}^{z} \lambda_{1}(t-z+u, u) \,\textrm du \right\}.
\label{k1}
\end{align}
Similarly, the probability density of being unmarried and out of school with $w$ years of schooling and alive aged $[z, z+\textrm dz)$ at time $t$, 
is proportional to $\beta_1(t-z, a_0) k_0(t, z, w) \,\textrm dz$ where 
\begin{align}
k_0(t,z,w) &= \exp\left\{-\int_{a_e}^{a_w} \alpha_{0}(t-z+u, u) \,\textrm du \right\} \nonumber \\
&\quad\times \alpha_{0}(t-z+a_w, a_w)^{\mathbf 1_{\{a_e < a_w \le z\}}} \nonumber \\ 
&\quad\times  \exp\left\{-\int_{a_e}^z \mu_{\mathbf 1_{\{u < a_w\}}0}(t - z + u, u) \,\textrm du \right\} \nonumber \\
&\quad\times \exp\left\{-\int_{a_0}^{z} \lambda_{\mathbf 1_{\{u \le a_w\}}}(t - z + u, u) \,\textrm du \right\},
\label{k0}
\end{align}
where $a_w = a_e + w < z,$ is the age when school ended with $w$ years of schooling attained.

Equations (\ref{k0}) and (\ref{k1}) can be combined and rewritten as follows.
\begin{align}
k(t,z,w) &= \exp\left\{-\int_{a_e}^{\min(a_w,z)} \alpha_{0}(t-z+u, u) \,\textrm du \right\} \nonumber \\
&\quad\times \alpha_{0}(t-z+a_w, a_w)^{\mathbf 1_{\{a_e < a_w \le z\}}} \nonumber \\ 
&\quad\times  \exp\left\{-\int_{a_e}^z \mu_{\mathbf 1_{\{u < a_w\}}0}(t - z + u, u) \,\textrm du \right\} \nonumber \\
&\quad\times \exp\left\{-\int_{a_0}^{z} \lambda_{\mathbf 1_{\{u < a_w\}}}(t - z + u, u) \,\textrm du \right\}.
\label{k}
\end{align}
Similarly, the probability density of being married and having $w$ years of schooling, alive and aged $[z, z+\textrm dz)$ at time $t$ and the first marriage at age $[y, y+ \textrm dy)$ is proportional to $\beta_1(t-z, a_0) h(t, y, z, w) \,\textrm dy \,\textrm dz$ where $h(t,y,z,w)$ is defined as
\begin{align}
h(t, y, z, w) &= \exp\left\{-\int_{a_e}^{\min(a_w,z)} \alpha_{\mathbf 1_{\{y < u\}}}(t-z+u, u) \,\textrm du \right\} \nonumber \\ 
&\quad\times \alpha_{\mathbf 1_{\{y < a_w\}}}(t-z+a_w,a_w)^{\mathbf 1_{\{a_e < a_w \le z\}}} \nonumber \\
&\quad\times \exp\left\{-\int_{a_e}^z \mu_{\mathbf 1_{\{u < a_w\}}\mathbf 1_{\{y < u\}}}(t - z + u, u) \,\textrm du \right\} \nonumber \\
&\quad\times \exp\left\{-\int_{a_0}^{y} \lambda_{\mathbf 1_{\{u < a_w\}}}(t - z + u, u)\,\textrm du\right\} \nonumber \\
&\quad\times \lambda_{\mathbf 1_{\{y < a_w\}}}(t-z+y,y). 
\label{h}
\end{align}
The likelihood contributions of individuals starting in state 3  (women who received no education) are defined similarly, but multiplied by $\beta_0(t-z,a_e)$, which is the probability density of being born in year $t-z$ and being in state 3 at age $a_e$, and taking $a_w = a_e$, in which case the $\alpha$ intensities do not appear in \eqref{k1}-\eqref{h}.

The probability density of the sampling event of being married, having $w$ years of schooling, alive and aged $[z, z+\textrm dz)$ at time $t$ is 
\begin{eqnarray*} 
\int_{a_0}^{z} \beta_{\mathbf 1_{\{a_e < a_w\}}}(t-z,a_e) h(t, y, z, w) \,\textrm dy.
\end{eqnarray*}
Alternatively, if we were interested in estimating intensities for both marriage and ending formal education, we could write the likelihood without conditioning on the education history. However, because our interest is in marriage intensity, we write the likelihood conditional on the education history, and consider conditions under which we can simplify the likelihood into a function of the marriage intensities alone.

The conditional likelihood contributions of individuals $i \in C_2$, in the prevalent cohort, e.g. NFHS-2, at time $t_2$ are
\begin{align*}
\prod_{i \in C_2} L_{2i}(\theta) &= \prod_{i \in C_2}  \frac{\beta_{\mathbf 1_{\{a_e < a_{w_i}\}}}(t_2-z_i,a_e) h(t_2, y_i, z_i, w_i)}{\int_{a_0}^{z_i} \beta_{\mathbf 1_{\{a_e < a_{w_i}\}}}(t_2-z_i,a_e) h(t_2, v, z_i, w_i) \,\textrm dv} \\
&= \prod_{i \in C_2}  \frac{h(t_2, y_i, z_i, w_i)}{\int_{a_0}^{z_i} h(t_2, v, z_i, w_i) \,\textrm dv}.
\end{align*}
The likelihood can be simplified under either of the following assumptions related to the counting process for number of schooling years, combined with the assumption that mortality is non-differential with respect to the marriage status, i.e. that $\mu_{j0}(t,a)=\mu_{j1}(t,a)=\mu_{j}(t,a)$ for $j = 0, 1$.
\begin{itemize}
\item[A1.] Schooling ends always before marriage or the intensity of stopping schooling after marriage is negligible.
\item[A2.] The intensities of stopping schooling are non-differential. That is the intensities of stopping schooling 
are the same before and after marriage, and does not depend on the history of the marriage process $\mathcal{F}_2(t, a) = \{N_2(s, u), u \le a, s(u) \le t(a)\}.$ In other words, this assumption states that the education process is locally independent of the marriage process; $\alpha_{0}(t,a\mid \mathcal{F}(a^-)) = \alpha_{1}(t,a\mid \mathcal{F}(a^-)) = \alpha(t,a\mid \mathcal{F}_1(a^-))$.  \citep{cook:2018}
\end{itemize}
\subsection{Retrospective cohort study I: likelihood under the assumptions of non-differential mortality and A2} 
Under the above-mentioned assumptions,  equation (\ref{h}) reduces to
\begin{align}
h(t, y, z, w) &= \exp\left\{-\int_{a_e}^{\min(a_w,z)} \alpha(t-z+u, u) \,\textrm du \right\} \nonumber \\ 
&\quad\times \alpha(t-z+a_w,a_w)^{\mathbf 1_{\{a_e < a_w \le z\}}} \nonumber \\
&\quad\times \exp\left\{-\int_{a_e}^z \mu_{\mathbf 1_{\{u < a_w\}}}(t - z + u, u) \,\textrm du \right\} \nonumber \\
&\quad\times \exp\left\{-\int_{a_0}^{y} \lambda_{\mathbf 1_{\{u < a_w\}}}(t - z + u, u)\,\textrm du\right\} \nonumber \\
&\quad\times \lambda_{\mathbf 1_{\{y < a_w\}}}(t-z+y,y). 
\end{align}
The normalising factor becomes
\begin{align}
\int_{a_0}^z h(t, y, z, w) \,\textrm dy &= \exp\left\{-\int_{a_e}^{\min(a_w,z)} \alpha(t-z+u, u) \,\textrm du \right\} \nonumber \\ 
&\quad\times \alpha(t-z+a_w,a_w)^{\mathbf 1_{\{a_e < a_w \le z\}}} \nonumber \\
&\quad\times \exp\left\{-\int_{a_e}^z \mu_{\mathbf 1_{\{u < a_w\}}}(t - z + u, u) \,\textrm du \right\} \nonumber \\
&\quad\times \left(1-\exp\left\{-\int_{a_0}^{z} \lambda_{\mathbf 1_{\{u < a_w\}}}(t - z + u, u)\,\textrm du\right\}\right). \nonumber \\
\label{hr}
\end{align}
Now the terms containing $\alpha$ and $\mu$ cancel out from the conditional likelihood, giving the likelihood contribution of an individual $i\in C_2$ in the prevalent cohort, e.g. NFHS-2, conditioned on the sampling event as
\begin{equation}\label{L2}
L_{2i}(\theta) = \frac{\exp\left\{-\int_{a_0}^{y_i} \lambda_{\mathbf 1_{\{u < a_{w_i}\}}}(t_2 - z_i + u, u)\,\textrm du\right\} \lambda_{\mathbf 1_{\{y_i < a_{w_i}\}}}(t_2-z_i+y_i,y_i)}{1 - \exp\left\{-\int_{a_0}^{z_i} \lambda_{\mathbf 1_{\{u < a_{w_i}\}}}(t_2 - z_i + u, u)\,\textrm du\right\}}.
\end{equation}

If we don't include the number of schooling years in the sampling event then the denominator will have to be integrated with respect to $w$ as well as
\begin{eqnarray*} 
\int_{a_0}^{z} \int_{a_e}^{z} \beta_{\mathbf 1_{\{a_e < a_w\}}}(t-z,a_0) h(t, y, z, w) \,\textrm dw \,\textrm dy.
\end{eqnarray*}
The above expression can be simplified under the assumptions A1 or A2 but the intensities $\alpha$ do not cancel out, so the resulting likelihood can be used for estimating parameters characterizing the education process, if these are of interest.

\subsection{Retrospective cohort study II: likelihood under the assumptions of non-differential mortality and A2} 
The conditional probability density of the sampling event of being alive with schooling years $w$  and aged $z$ at time $t$  is the sum of the probabilities of being (unmarried, alive with schooling years $w$ and aged $z$ at $t$), and (married, alive with schooling years $w$ and aged $z$ at $t$). This is given by $\beta_{\mathbf 1_{\{a_e < a_w\}}}(t-z, a_e) [k(t,z,w) + \int_{a_0}^z h(t, y, z, w) \,\textrm dy]$. Thus, the conditional likelihood contributions of individuals $i \in C_3$ in the general cohort, e.g.  NFHS-3, at time $t_{3}$ are
\begin{align}\label{L3i}
L_3(\theta) &= \prod_{i \in C_3}  \frac{\beta_{\mathbf 1_{\{a_e < a_{w_i}\}}}(t_{3}-z_i,a_0) h(t_3, y_i, z_i, w_i)^{\delta_i} k(t_{3}, z_i, w_i)^{1 - \delta_i}}{\beta_{\mathbf 1_{\{a_e < a_{w_i}\}}}(t_{3}-z_i,a_0) [k(t_3, z_i, w_i) + \int_{a_0}^{z_i} h(t_{3}, u, z_i, w_i) \,\textrm du]},
\end{align}
where $\delta_i \equiv \mathbf 1_{\{y_i \le z_i\}}$ is an indicator of marital status at time $t_3$. Under the assumptions A2 and non-differential mortality with respect to marriage status, as before, \eqref{k} reduces to
\begin{align}
k(t,z,w) &= \exp\left\{-\int_{a_e}^{\min(a_w,z)} \alpha(t-z+u, u) \,\textrm du \right\} \nonumber \\
&\quad\times \alpha(t-z+a_w, a_w)^{\mathbf 1_{\{a_e < a_w \le z\}}} \nonumber \\ 
&\quad\times  \exp\left\{-\int_{a_e}^z \mu_{\mathbf 1_{\{u < a_w\}}}(t - z + u, u) \,\textrm du \right\} \nonumber \\
&\quad\times \exp\left\{-\int_{a_0}^{z} \lambda_{\mathbf 1_{\{u < a_w\}}}(t - z + u, u) \,\textrm du \right\}
\label{kr}
\end{align}
and $h(t, u, z, w)$ to \eqref{hr}. Combining these, the normalising factor becomes
\begin{align}
\MoveEqLeft k(t, z, w) + \int_{a_0}^{z} h(t, u, z, w) \,\textrm du \nonumber \\
&= \exp\left\{-\int_{a_e}^{\min(a_w,z)} \alpha(t-z+u, u) \,\textrm du \right\} \alpha(t-z+a_w, a_w)^{\mathbf 1_{\{a_e < a_w \le z\}}} \nonumber \\ 
&\quad\times  \exp\left\{-\int_{a_e}^z \mu_{\mathbf 1_{\{u < a_w\}}}(t - z + u, u) \,\textrm du \right\},
\end{align}
which will cancel out with the similar term in the numerator of the conditional likelihood. Thus, under these assumptions, the likelihood \eqref{L3i} under the retrospective cross-sectional design II reduces to the standard likelihood for right censored survival data, given by 
\begin{align}\label{L3}
L_3(\theta)  &= \prod_{i \in C_3} \bigg[ \lambda_{\mathbf 1_{\{y_i < a_w\}}}(t_3 - z_i + y_i, y_i)^{\delta_i} \nonumber \\
&\qquad\times \exp\left\{-\int_{a_0}^{\min(y_i,z_i)} \lambda_{\mathbf 1_{\{u < a_w\}}}(t - z + u, u) \,\textrm du\right\} \bigg].
\end{align}
It is to be noted that the above likelihood expressions are constructed by explicitly conditioning on the calendar time of the survey, the age and schooling status  of the individual at the  time of the survey. This is equivalent to conditioning on the individual's birth cohort, and hence, the birth rate cancels out and the likelihood expressions simplify by assuming non-differential mortality and either $A_1$ or $A_2$ only. If the conditional likelihood were derived by conditioning only on the age range used for the sampling and not on the exact age of individuals then the probability of the conditioning event needed to be integrated over $z$ also. The same applies for education status. In this case, stricter assumptions would be needed to carry out the estimation of the incidence rate or external information on mortality, education as well as birth rates would be required. Such information may not be available for all the stratifying groups that we will use in the real application. In the following we use a likelihood conditioned on the covariates and the sampling scheme for estimating the marriage incidence rate. The likelihood is a product of $L_2(\theta)$ and $L_3(\theta)$ from the cohorts under design I and II, respectively. We show in Appendix B that this is indeed a likelihood and hence, the maximum likelihood theory applies for estimation of $\theta$.
\section{Predictive probabilities}\label{section:predictive}
Given characteristics $x$, we might be interested in the predictive probability of an unmarried woman aged $a_1$ $(\ge a_0)$ at time $t$ and schooling years $w$ years being married before age $a_2$. Because education is time-dependent, generally calculation of these kinds of probabilities would involve prediction of future education also. However, for women who already reached their highest level of education (i.e. $a_w < a_1$), we can predict based on marriage intensity and mortality estimates alone. Such predictive probability for fixed schooling years is given by the cumulative incidence
\begin{align}\label{equation:predictive}
\MoveEqLeft \text{PredProb}(a_2\mid t, a_1, w) \nonumber \\
&=   \frac{\int_{a_1}^{a_2} k(t -a_1 +a, a, w) \lambda_0(t - a_1 + a, a;\theta) \,\textrm da}{k(t, a_1,w)} \nonumber \\
&= \frac{\int_{a_1}^{a_2} k_2(t -a_1 +a, a, w) \lambda_0(t - a_1 + a, a;\theta) \,\textrm da}{k_2(t, a_1,w)},
\end{align}
where 
\begin{align*}
\MoveEqLeft k_2(s, a, w) \\ 
&= \exp\left\{-\int_{a_0}^{a} [\mu_{00}(s-a+u, u) + \lambda_0(s -a +u, u;\theta)\}]\,\textrm du \right\}.
\end{align*}
Another predictive probability of interest is that of an unmarried woman, with characteristics $x$ and aged $a_1$ at time $t$ and schooling years $w_1$
being married before age $a_2$, and being alive at $a_2$, and is given by (for fixed schooling years)
\begin{align}\label{equation:predictive2}
\MoveEqLeft \frac{\int_{a_1}^{a_2} h(t - a_1 + a_2, a, a_2, w_1) \,\textrm da}{k(t,a_1,w_1)} \nonumber \\
&= \frac{\int_{a_1}^{a_2} h_2(t - a_1 + a_2, a, a_2, w_1) \,\textrm da}{k_2(t,a_1,w_1)}, 
\end{align}
where $k_2$ is defined above and $h_2$ is obtained from $h$ in \eqref{h} by dropping terms corresponding to education process. Under the assumption of non-differential mortality with respect to both education and marriage, and possibly other covariates used to model marriage intensity, mortality rates based on official statistics can be used in the calculation.

The first predictive probability (\ref{equation:predictive}) appears to be important for population models since it gives the proportion ever getting married,
which multiplied by the population count of age $a_1$ (with characteristics $x$) gives the ever-married population count.
The second one (\ref{equation:predictive2}) might be important for questions like: what proportion of women of age $a_0$ get married and live until through a typical 
``child-bearing age'' $a_1$. Note that the mortality rate is needed in order to compute above probabilities. We demonstrate the former kind of predictive probabilities in Section \ref{section:results}.

\section{Simulation study}\label{section:simulation}
We conducted a simulation study to assess the efficiency gain achieved by combining data from the two retrospective cohort studies, compared to analysing each one of these separately, as well as to study the impact of various misspecification scenarios on the parameter estimates. We simulated data from a multi-state model with states similar to Figure~\ref{figure:multistate}. The model was specified through the transition rates in the transition intensity matrix
\[
\begin{blockarray}{c@{\hspace{1pt}}rrrrr@{\hspace{4pt}}}
 & 1   & 2   & 3 & 4 & 5\\
\begin{block}{r@{\hspace{1pt}}|@{\hspace{1pt}}@{\hspace{1pt}}rrrrr@{\hspace{1pt}}@{\hspace{1pt}}|}
1 & . & \lambda_1 = e^{m + bx + c} & \alpha_0 = e^{s + bx} & 0 & \mu_{10} = e^{r + bx + c}\\
2 & 0 & . & 0 & \alpha_1 = e^{s + bx + d} & \mu_{11} = e^{r + bx + c + g} \\
3 & 0 & 0 & . & \lambda_0 = e^{m + bx} & \mu_{00} = e^{r + bx}\\
4 & 0 &  0 & 0  & . & \mu_{01}  = e^{r + bx + g} \\
5 & 0 &  0 & 0  & 0 & .  \\
\end{block}
\end{blockarray}\]
where the parameters of interest were $m$, characterising the baseline marriage rate, $b$, characterising the effect of a time constant covariate $x$ (taking values 1 or 0 with probability 0.5) on marriage rate, and $c$, characterizing the effect of being in school on marriage rate, as well as on mortality rate. Parameters $d$ and $g$ characterise the effect of marriage on ending formal education and mortality, respectively. Note that  $d=g=0$ under the non-differential assumption,. The initial state of the multi-state model at age $a_0=12$ was drawn randomly with probabilities $\textrm{expit}(-1+0.5 x)$ for state 3 (unmarried, out of school) and $1-\textrm{expit}(-1+0.5 x)$ for state 1 (unmarried, in school). A cross-sectional cohort was constructed by drawing year of birth uniformly from $[1965,1993]$, and taking 2005 as the time of the cross-sectional survey, at which time the age range was $[12,40]$. The cohort under design I (cohort I) was constructed by simulating 2,500 event histories and including only individuals in the married states 2 and 4 at the time of the cross-section, while the one under design II (cohort II) was constructed by simulating 2,500 event histories and including individuals in states 1-4 at the cross-section. To these data, we fitted constant rate marriage models through maximizing the joint likelihood expression \eqref{L2} \& \eqref{L3} (both cohorts), \eqref{L2} (cohort I data only), \eqref{L2} without the correction term in the denominator, and \eqref{L3} (cohort II data only). The data generation and model fitting were repeated 1,000 times, resulting in average sizes of the two cohorts as (1,864, 2,229), respectively under the non-differential scenario. The likelihood expressions were maximised numerically using the R \texttt{optim} function \citep{rmanual}, with standard errors calculated by inverting the numerically differentiated Hessian matrix at the maximum likelihood point.
\\
The results under the non-differential scenario are given in Table \ref{table:simtable1}. The results indicate that there is a clear efficiency gain (in terms of the Monte Carlo standard deviation of the point estimates) in combining the analysis of the two cohorts, as opposed to analysing each of them separately. The three types of parameters, baseline marriage intensity $m$, effect of a time-constant covariate $b$, and effect of time-dependent covariate $c$ can be estimated without bias, with the cohort I likelihood needing the correction term in the denominator to account for the sampling mechanism. The results under the scenario of differential education process intensities are given in Table \ref{table:simtable2}. These indicate that violation of the non-differential assumption for stopping school  mainly causes bias in the estimated effect of ending school on marriage incidence, while the other two parameters are much less affected. The retrospective cohort likelihood under design I is more susceptible to this type of bias, but it is fairly small in all cases. Differential mortality (Table \ref{table:simtable3}) on the other hand causes bias in the baseline marriage incidence estimates with both types of likelihood expression, with the covariate effect estimates affected much less. Finally, both types of non-differential assumptions are combined in the scenario of Table \ref{table:simtable4}, with the two different types of biases essentially adding up. In summary, the simulation results confirm the efficiency gain in combining two types of retrospective cross-sectional cohort data, while demonstrating how the parameter estimates are affected by violations of the assumptions used in deriving the simplified likelihood expressions. Because the effect of the violations was relatively small, we proceed under the non-differential assumptions in the data analysis of Section \ref{section:results}.

\begin{table}[!h]
\small\sf\centering
\caption{Results from 1000 simulation rounds under non-differential mortality and stopping school ($d=g=0$). Mean stands for mean point estimate, MC SD for Monte Carlo standard deviation of the point estimates, Mean SE for mean estimated standard error, and Coverage for 95\% confidence interval coverage probability.}
\centering
\begin{tabular}{ccrrrrrr}
\hline
Likelihood & Parameter & Truth & Mean & Bias & MC SD & Mean SE & Coverage \\
\hline
\eqref{L2} \& \eqref{L3} & $m$ & -1.500 & -1.501 & -0.001 & 0.029 & 0.030 & 0.947 \\
 & $b$ & 0.500 & 0.501 & 0.001 & 0.037 & 0.037 & 0.951 \\
 & $c$ & -0.500 & -0.499 & 0.001 & 0.038 & 0.039 & 0.943 \\
\eqref{L2} & $m$ & -1.500 & -1.501 & -0.001 & 0.049 & 0.050 & 0.954 \\
 & $b$ & 0.500 & 0.501 & 0.001 & 0.059 & 0.060 & 0.942 \\
 & $c$ & -0.500 & -0.499 & 0.001 & 0.059 & 0.058 & 0.943 \\
\eqref{L2} & $m$ & -1.500 & -1.264 & 0.236 & 0.034 & 0.037 & 0.000 \\
w/o & $b$ & 0.500 & 0.394 & -0.106 & 0.044 & 0.046 & 0.380 \\
correction & $c$ & -0.500 & -0.474 & 0.026 & 0.052 & 0.052 & 0.901 \\
\eqref{L3} & $m$ & -1.500 & -1.501 & -0.001 & 0.038 & 0.037 & 0.941 \\
 & $b$ & 0.500 & 0.501 & 0.001 & 0.047 & 0.046 & 0.947 \\
 & $c$ & -0.500 & -0.500 & 0.000 & 0.052 & 0.052 & 0.947 \\

\hline
\end{tabular}
\label{table:simtable1}
\end{table} 

\begin{table}[!h]
\small\sf\centering
\caption{Results from 1000 simulation rounds under non-differential mortality ($g=0$) and differential stopping school ($d=1$).}
\centering
\begin{tabular}{ccrrrrrr}
\hline
Likelihood & Parameter & Truth & Mean & Bias & MC SD & Mean SE & Coverage \\
\hline
\eqref{L2} \& \eqref{L3} & $m$ & -1.500 & -1.507 & -0.007 & 0.030 & 0.030 & 0.948 \\
 & $b$ & 0.500 & 0.502 & 0.002 & 0.036 & 0.037 & 0.951 \\
 & $c$ & -0.500 & -0.470 & 0.030 & 0.038 & 0.038 & 0.864 \\
\eqref{L2} & $m$ & -1.500 & -1.516 & -0.016 & 0.052 & 0.051 & 0.944 \\
 & $b$ & 0.500 & 0.501 & 0.001 & 0.060 & 0.060 & 0.953 \\
 & $c$ & -0.500 & -0.429 & 0.071 & 0.057 & 0.057 & 0.752 \\
\eqref{L2} & $m$ & -1.500 & -1.264 & 0.236 & 0.035 & 0.037 & 0.000 \\
w/o & $b$ & 0.500 & 0.395 & -0.105 & 0.045 & 0.046 & 0.370 \\
correction & $c$ & -0.500 & -0.478 & 0.022 & 0.051 & 0.052 & 0.932 \\
\eqref{L3} & $m$ & -1.500 & -1.500 & -0.000 & 0.038 & 0.037 & 0.935 \\
 & $b$ & 0.500 & 0.502 & 0.002 & 0.048 & 0.046 & 0.948 \\
 & $c$ & -0.500 & -0.505 & -0.005 & 0.052 & 0.052 & 0.947 \\

\hline
\end{tabular}
\label{table:simtable2}
\end{table} 

\begin{table}[!h]
\small\sf\centering
\caption{Results from 1000 simulation rounds under differential mortality ($g=1$) and non-differential stopping school ($d=0$).}
\centering
\begin{tabular}{ccrrrrrr}
\hline
Likelihood & Parameter & Truth & Mean & Bias & MC SD & Mean SE & Coverage \\
\hline
\eqref{L2} \& \eqref{L3} & $m$ & -1.500 & -1.547 & -0.047 & 0.032 & 0.032 & 0.709 \\
 & $b$ & 0.500 & 0.498 & -0.002 & 0.040 & 0.041 & 0.954 \\
 & $c$ & -0.500 & -0.497 & 0.003 & 0.042 & 0.042 & 0.945 \\
\eqref{L2} & $m$ & -1.500 & -1.557 & -0.057 & 0.058 & 0.056 & 0.831 \\
 & $b$ & 0.500 & 0.501 & 0.001 & 0.070 & 0.069 & 0.946 \\
 & $c$ & -0.500 & -0.496 & 0.004 & 0.067 & 0.065 & 0.943 \\
\eqref{L2} & $m$ & -1.500 & -1.283 & 0.217 & 0.038 & 0.039 & 0.001 \\
w/o & $b$ & 0.500 & 0.393 & -0.107 & 0.050 & 0.051 & 0.443 \\
correction & $c$ & -0.500 & -0.467 & 0.033 & 0.057 & 0.056 & 0.904 \\
\eqref{L3} & $m$ & -1.500 & -1.543 & -0.043 & 0.039 & 0.039 & 0.811 \\
 & $b$ & 0.500 & 0.497 & -0.003 & 0.050 & 0.051 & 0.951 \\
 & $c$ & -0.500 & -0.499 & 0.001 & 0.058 & 0.056 & 0.938 \\

\hline
\end{tabular}
\label{table:simtable3}
\end{table} 

\begin{table}[!h]
\small\sf\centering
\caption{Results from 1000 simulation rounds under differential mortality and stopping school ($g=d=1$).}
\centering
\begin{tabular}{ccrrrrrr}
\hline
Likelihood & Parameter & Truth & Mean & Bias & MC SD & Mean SE & Coverage \\
\hline
\eqref{L2} \& \eqref{L3} & $m$ & -1.500 & -1.555 & -0.055 & 0.032 & 0.032 & 0.595 \\
 & $b$ & 0.500 & 0.498 & -0.002 & 0.039 & 0.041 & 0.962 \\
 & $c$ & -0.500 & -0.469 & 0.031 & 0.043 & 0.042 & 0.874 \\
\eqref{L2} & $m$ & -1.500 & -1.577 & -0.077 & 0.058 & 0.057 & 0.741 \\
 & $b$ & 0.500 & 0.504 & 0.004 & 0.068 & 0.069 & 0.952 \\
 & $c$ & -0.500 & -0.419 & 0.081 & 0.065 & 0.063 & 0.748 \\
\eqref{L2} & $m$ & -1.500 & -1.282 & 0.218 & 0.038 & 0.039 & 0.000 \\
w/o & $b$ & 0.500 & 0.394 & -0.106 & 0.049 & 0.051 & 0.472 \\
correction & $c$ & -0.500 & -0.478 & 0.022 & 0.057 & 0.056 & 0.921 \\
\eqref{L3} & $m$ & -1.500 & -1.542 & -0.042 & 0.040 & 0.039 & 0.809 \\
 & $b$ & 0.500 & 0.496 & -0.004 & 0.052 & 0.051 & 0.944 \\
 & $c$ & -0.500 & -0.509 & -0.009 & 0.057 & 0.056 & 0.947 \\

\hline
\end{tabular}
\label{table:simtable4}
\end{table}

\vspace{.1in} 
\section{Analysis of marriage incidence in India and four states based on NFHS-2 and -3 data}\label{section:results}
\subsection{Model}\label{section:model}
Our main focus is on estimation of marriage incidence and hence, we apply estimation method based on likelihood expressions (\ref{L2}) and (\ref{L3}). The NFHS data did not include complete retrospective histories of education and hence, we constructed histories of education process up to the time of marriage by employing the structure of the Indian education system and assuming that everyone starts school at the same age, and stays at school continuously after that until stopping. 

For the purpose of illustration, we considered only four states, Kerala, Maharashtra, Punjab and Rajasthan, as described in Section~\ref{section:data} which are geographically spread across India and differ by way of literacy rates, women's position, and sex-ratio at birth.  
From the cross-sectional data from NFHS-2 and NFHS-3 surveys, the age and calendar time effects turned out to be strongly correlated, so instead of estimating age- and calendar period- specific marriage incidence rates, we used age as the main time scale of the analysis, and birth cohort as a covariate. Table~\ref{table:covariates} lists the covariates used in the marriage incidence model. We applied a proportional hazards model in which the covariates act multiplicatively on an age-dependent baseline rate, assumed piecewise constant over one-year age intervals except for the first and the last intervals. This results in 17 age bands $[12,15), [15,16), \ldots, [29,30)$, and $[30,50)$ years, denoted by $[a_j,a_{j+1})$, $j=1,\ldots,17$. 

 The effect of education was modeled by defining the marriage incidence as a function of the current highest education level being attempted, defining the education level $x_{5j}$ at age band $j$ as a time-dependent covariate by modifying the woman's highest attained level $x_{5}$, recorded at the time of survey, so that $x_{5j} = \min(x_{5}, 1)$ when in age band $j=1$, $x_{5j} = \min(x_{5}, 2)$ when $j \in \{2,3,4\}$, and $x_{5j} = x_{5}$ otherwise. For example, consider a woman aged 25 years at the time of survey, married at the age of 21 years, has reported education level \emph{Secondary} ($x_5 = 2$ (cf. Figure~\ref{figure:samplepath} and Table~\ref{table:covariates}). In the analysis, her contribution to the education variable will be \emph{Primary} in the age band $[12,15)$ and \emph{Secondary} in the bands $[15,16)$, $[16,17)$, $\cdots$, $[20,21)$ and $[21,22)$ to span the age range from 12 years to the age at her marriage. 

To sum up, the model for the marriage incidence rate $\lambda(a; x, \theta) = \lambda_{j}(x, \theta)$, $a \in [a_j, a_{j+1})$, $j=1,\ldots, 17,$ conditional on covariate values $x$ was specified as
\begin{align}\label{equation:incidencemodell}
\log\{\lambda_{j}(x, \theta)\} &= \alpha_{j} + \sum_{i=1}^3 \beta_{1i} \mathbf 1_{\{x_1 = i\}}
+ \beta_2 \mathbf 1_{\{x_2 = 1\}}
 + \sum_{i=1}^3 \beta_{3i} \mathbf 1_{\{x_3 = i\}} \nonumber \\
&\quad + \sum_{i=1}^4 \beta_{4i} \mathbf 1_{\{x_4 = i\}}
 + \sum_{i=1}^3 \beta_{5i} \mathbf 1_{\{x_{5j} = i\}}, 
\end{align}
with 31 parameters, including 17 log-baseline rates and 14 covariate effects (log-rate ratios). The same model was fitted for each state separately, and in addition to all-India data (all 29 states) to assess how the state-specific patterns differ from the national pattern, by maximising the product of likelihood expressions of the form \eqref{L2} and \eqref{L3}, but because the age at marriage was only reported at the precision of one year, expressing the numerator contribution for married women as
\begin{align*}
\MoveEqLeft \int_{\floor*{y}}^{\ceil*{y}} \lambda(a; x, \theta) \exp\left\{-\int_{a_0}^{a} \lambda(u; x, \theta) \,\textrm du\right\} \,\textrm da \\
&= \exp\left\{-\int_{a_0}^{\floor*{y}} \lambda(u; x, \theta) \,\textrm du\right\} \left[1-\exp\left\{-\int_{\floor*{y}}^{\ceil*{y}} \lambda(u; x, \theta) \,\textrm du\right\} \right]
\end{align*}
where $\floor*{y}$ and $\ceil*{y} = \floor*{y} + 1$ denote the floor and ceiling of the exact age $y$ at which the marriage took place. The joint likelihood expression was maximised with respect to the parameter vector $\theta$ using the \texttt{optim} function of the R statistical environment \citep{rmanual}.  The standard errors were evaluated by inverting the numerically differentiated observed information matrix at the maximum likelihood point. The results were presented as point estimates and 95$\%$ confidence intervals. Of note, by letting the marriage rate depend on the birth cohort, the third possible time scale (calendar time) can be omitted. 

\subsection{Results}

\begin{figure}[!ht]
\centerline{\includegraphics[width=0.85\textwidth]{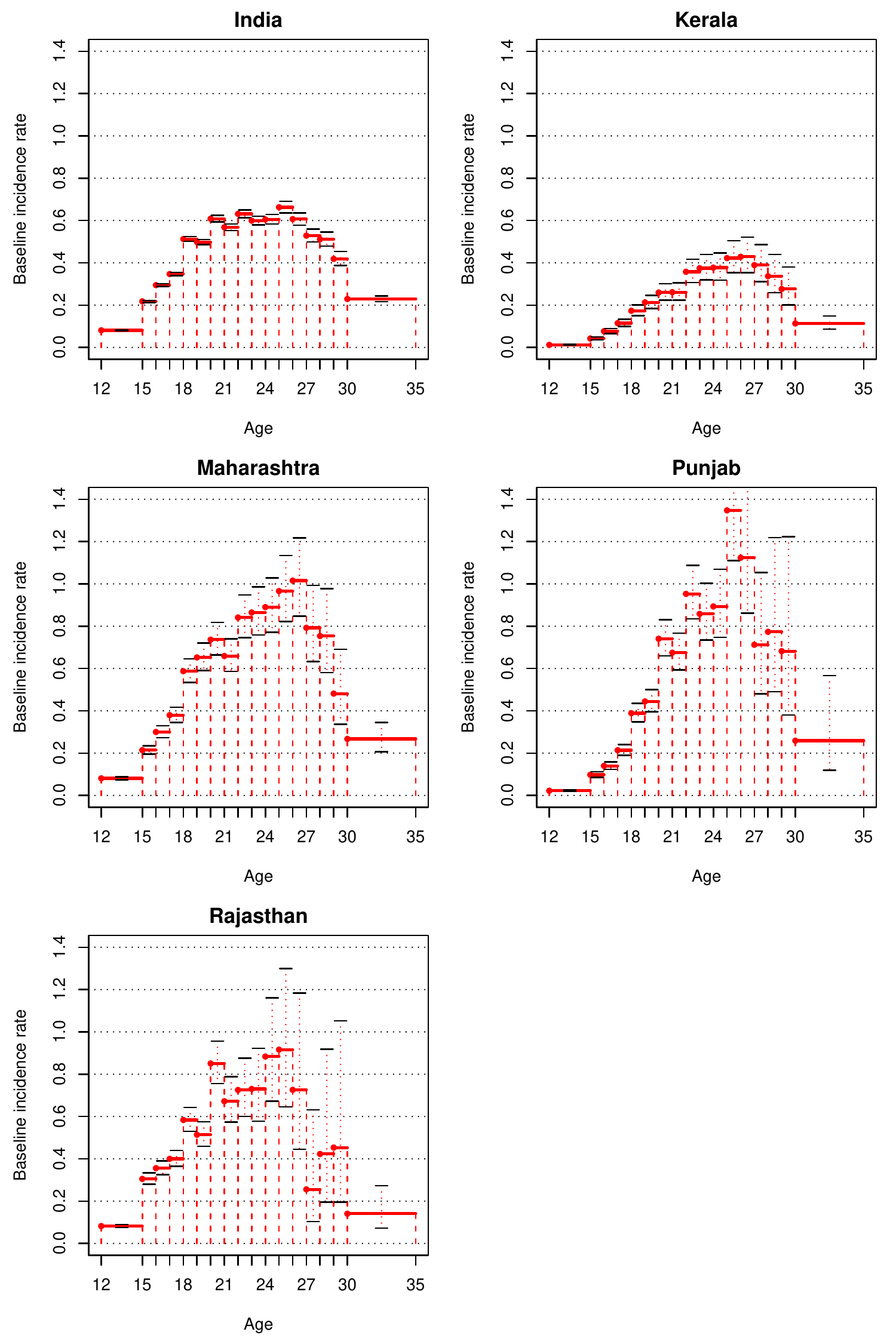}}
\caption{Age-specific baseline rates for a woman to marry in India and the four selected states. The horizontal lines
show the maximum likelihood estimates of parameters $\exp\{\alpha_{j}\}$ in \eqref{equation:incidencemodell}, 
and their corresponding 95\% confidence intervals.}\label{figure:baselinerates}
\end{figure}

Figure~\ref{figure:baselinerates}  presents the estimated age-specific baseline marriage rates in the four Indian states and in all India. Although the hazard of first marriage after age 30 has remained low in each state, different patterns emerge otherwise. The rate is generally lowest in Kerala, in particular in comparison to Maharashtra and Punjab. In Rajasthan, the rate starts increasing earliest in age.

\begin{figure}[!ht]
\centerline{\includegraphics[width=0.85\textwidth]{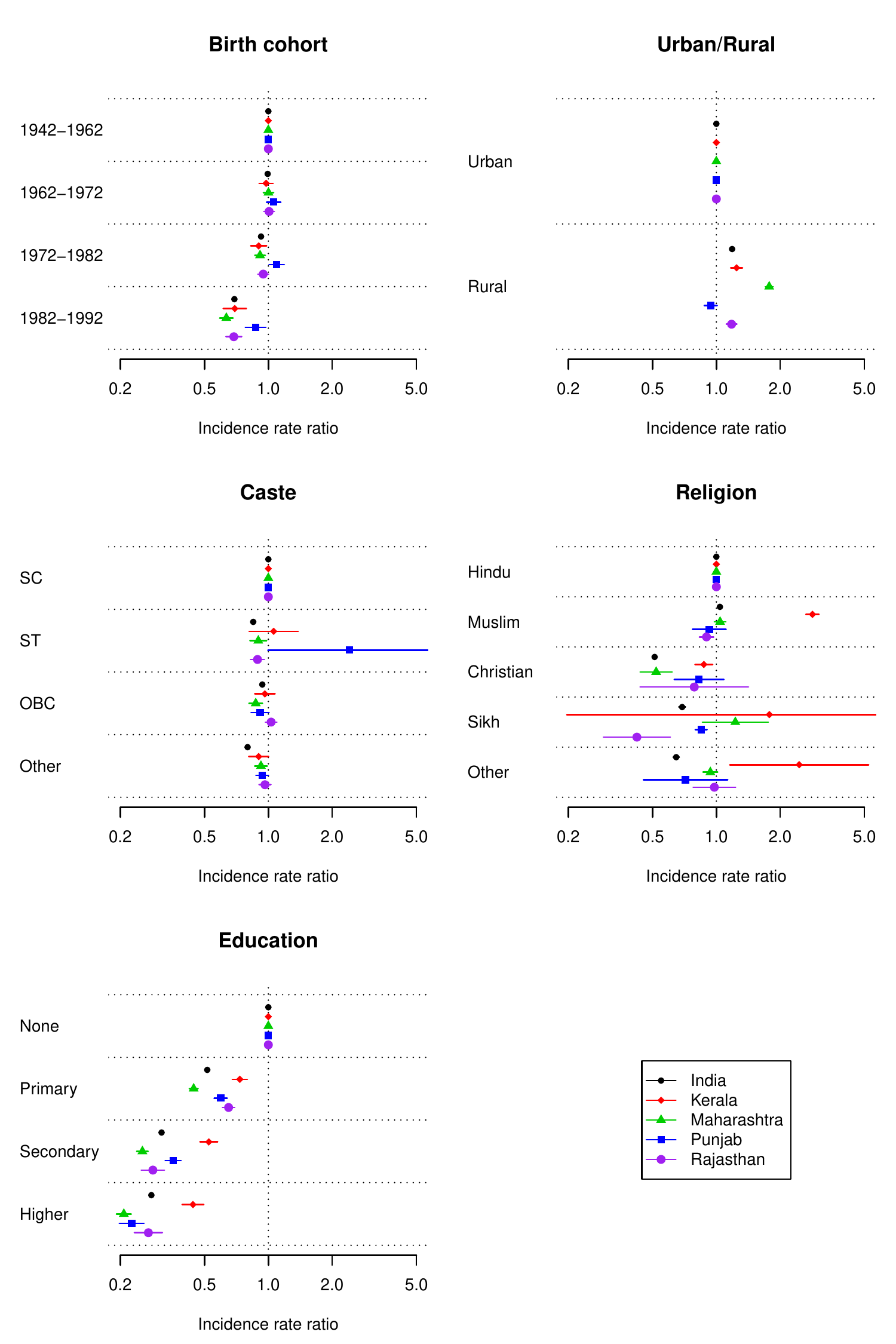}}
\caption{Forest plots of the estimated covariate effects on marriage incidence rates of women for India and the four selected states. 
The horizontal lines correspond to the rate ratio estimate, and 95\% confidence interval.}\label{figure:incidencemodelcovariates}
\end{figure}

Figure~\ref{figure:incidencemodelcovariates} shows the estimated covariate effects on the marriage rates. The rate decreases by birth cohort, except for Punjab where the rate is the highest for the 1972-1982 cohort. By the last cohort in this analysis (1982-1992), the rates have declined considerably in all four states. Since this birth cohort, being 6-16 years of age at the time of survey, was underrepresented in NFHS-2, we repeated the analysis by using only the NFHS-3 data and the estimates of marriage rates were essentially unchanged (results not shown).

Unsurprisingly, women in rural areas have a larger rate of marriage (all India incidence rate ratio of $1.19$) compared to urban areas, except in Punjab where the reverse is true. The higher rate in rural areas is particularly striking in Kerala and Maharashtra. At the India level, the marriage rates are similar for OBC and SC while ST and Other caste have lower marriage rates. However, this pattern is not evident in all of the four the state-level results. In Punjab, the confidence interval for ST is wide because this caste is rare (Table~\ref{table:descriptivewomen}).

There are clear differences in the marriage rate across religions. At the India level, the marriage incidence rates are clearly smaller in Christian, Sikh and other religions as compared to Hindu. The same pattern emerges in the state-level analysis, except for Muslims in Kerala. Again, to interpret the state specific results we note that not all religions were sufficiently represented in each state (Table~\ref{table:descriptivewomen}).

The effect of education is evident.  There is a clear decrease in the incidence rate when moving from no education to higher education levels in India and in all the four states. In the all India analysis, the incidence rate for a woman with primary education to marry at any given age is about half that for a woman with no education. The corresponding rates are 31$\%$ and 28$\%$ of the uneducated rate for a woman with secondary and higher education. The same patterns shows up in all four states although the effect of education level is relatively smaller in Kerala. 

\begin{figure}[!ht]
\centerline{\includegraphics[width=0.85\textwidth]{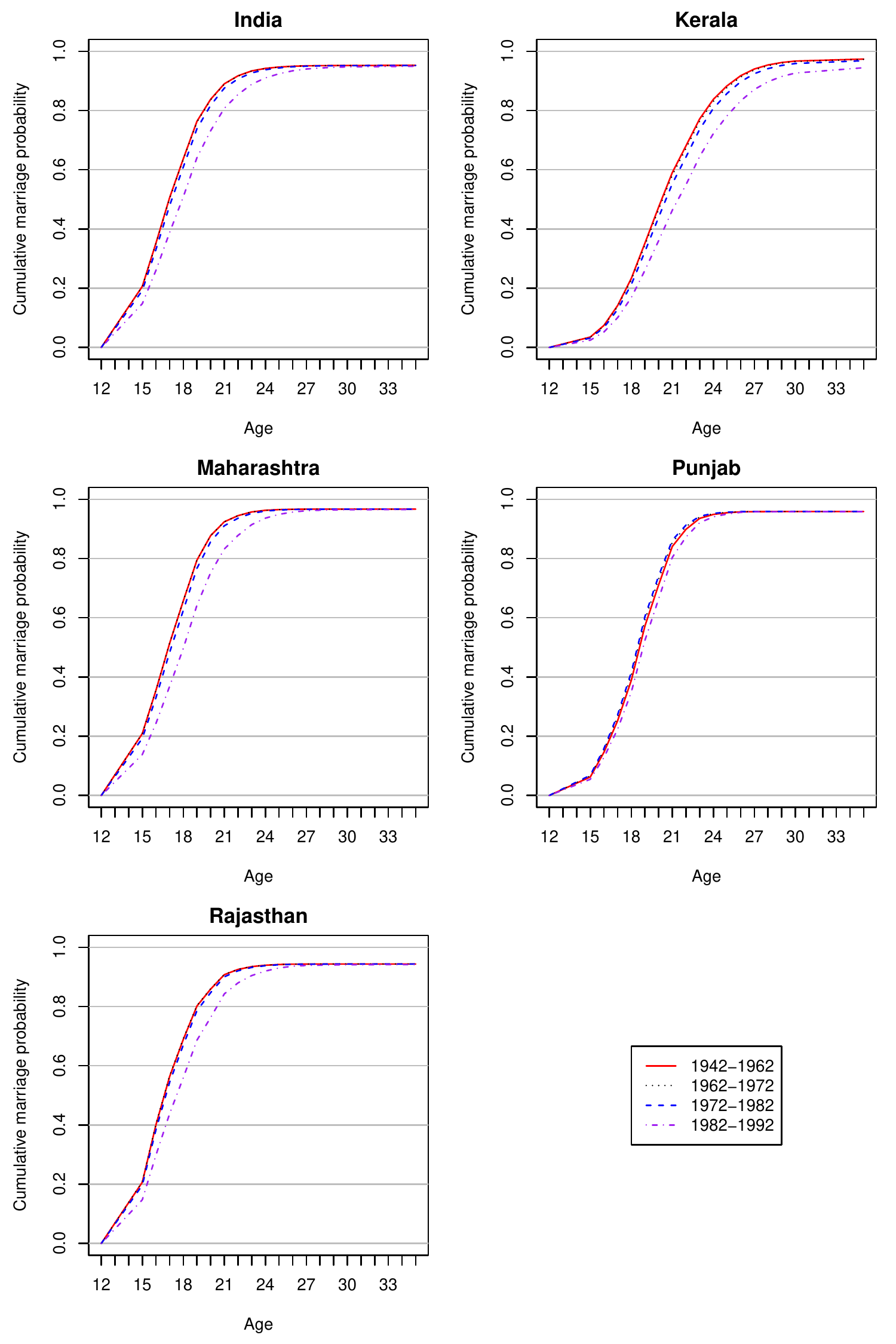}}
\caption{Predictive probabilities for women to be married by age $a$ by birth cohort, calculated by combining 2010 mortality rates with the marriage incidence model. The other covariates were set to the reference levels.}\label{figure:predictive}
\end{figure}

Predictive probabilities of type \eqref{equation:predictive} for marrying by age $a_1$ were calculated as discussed in Section \ref{section:predictive}, with $a_{\min} = a_0 = 12$, using 2010 mortality rates based on census data, and marriage incidence rates corresponding to different calendar periods (Figure \ref{figure:predictive}). The covariate values were set to the reference categories (urban area, scheduled caste, Hindu religion, and uneducated). Clearly, the women's absolute probability of marrying by late twenties has remained consistently high, but in Maharashtra there has been a clear shift towards marrying at a later in life. The patterns in Kerala and Rajasthan are more difficult to interpret, as the high estimated marriage rates in late twenties in the later calendar periods actually results also in higher projected absolute probabilities in late twenties. However, this projection does not reflect all the changes in the background population, since the overall education level has increased over time, bringing the population marriage incidence rates down, while in this projection education was fixed to the reference level.
In Punjab, any changes over time have been comparatively small.

\vspace{.1in} 
\section{Discussion}\label{section:discussion}
In this article we formulated a multi-state model for modeling an outcome and a covariate process jointly in two types of retrospective cross-sectional cohort studies. Our methodological contributions can be summarised as follows. (i) Combined analysis of retrospective histories from two types of cross-sectional cohorts; (ii) multi-state modeling of retrospective histories of two correlated processes in two time scales;  (iii) assessment of the performance of the method based on combining the two retrospective cross-sectional cohort designs against using either of the two; and (iv) illustration through an application to the estimation of marriage incidence rates. We have used explicitly the structure of the Indian education system in building the joint model and also extracting retrospective information from the cross-section. We also assume  that everyone adhere to that. When retrospective history on schooling is available, in addition to the cross-section, this assumption can be relaxed.

Statistical methods have been developed and applied for the estimation of incidence rates from cross-sectional cohorts, with or without subsequent prospective follow-up \citep{keiding:1991, keiding:2006, saarela:2009, keiding:2012}. The incidence rate, in general, is not identifiable from data under retrospective cross-section desing I only without supplementary information, e.g., data from design II. The estimation is simplified under assumptions such as time homogeneity and non-differential mortality before and after the incident event \citep{keiding:1991}. Much of the existing literature has focused on nonparametric estimation of cumulative incidence and survival functions through appropriately weighting the risk sets. Herein our main focus was in factors that modify the incidence rates, and therefore we applied likelihood-based methods for piecewise constant hazard models. For this purpose, we needed to combine likelihood functions arising from two different sampling plans, namely the cross-sectional cohort setting of NFHS-2, and the setting of NFHS-3. To combine information collected under the different sampling plans, the likelihood contributions from the individual surveys are conditioned on the specific sampling plan employed in the survey, with the overall likelihood expression obtained simply as the product of these. 

This result was applied in the estimation of the marriage incidence rates in four Indian states as functions of age and birth cohort, as well as demographic characteristics. Unlike previous approaches \citep{kashyap:2015} to estimate marriage rates, the proposed method allows combining information from more than one survey and modelling education and marriage jointly. This brings several advantages. First, the increased sample size leads to more powerful analyses of age at marriage data at the sub-population level (e.g. Indian states).  Second, it also allows learning of calendar time trends in the strength of association of many factors affecting marriage rates.

The analysis goes beyond simply describing the age- and sex-based marriage rates and puts forward a model which takes into account the well-recognised factors driving the marriages in India. The marriage incidence rates differ regionally (or state-wise) and hence the rates obtained using the India-level data may not bring out the real marriage squeeze problem existent in social strata defined by caste, religion and education. Although the caste effect on the marriage incidence rates did not differ much by state, those of education and religion did. Our analysis provides strong evidence towards religion, education and urban/rural area as the main factors affecting the marriage pattern among women in India. Education levels or qualifications seem to be replacing the earlier role of caste in shaping the marriage market in India. The effects of women's educational expansion on marriage incidence have been studied worldwide and found to have some impact. However, a considerable portion of the reduction in early marriage is not explained by changes in levels of education \citep{mensch:2005}. To predict the real magnitude of the marriage squeeze problem in India, predictions of married and unmarried populations in different age and social strata defined by state, caste, religion, urban/rural, and education are needed. The model proposed here will have a direct application for such predictions.

\section*{Acknowledgements}

The first two authors were partly supported by the project `Precarious family formation' financed by the Kone foundation. The first author's work was also supported by the research mobility grant (No. 325990) awarded by the Academy of Finland.

\clearpage

\vspace{.1in} 

\section*{Appendices}

\subsection*{Appendix A: Data selection and description}
The NFHS reports clearly bring out differences between the states with respect to education (\url{http://rchiips.org/nfhs/}). All four states considered here show increasing trends in the proportion of women attaining higher education but differ by education attainment. There is a decreasing trend in the proportion of 
primary and no education, and increasing trend in the secondary and higher education level. Rajasthan stands out when looking the education levels of women, with the highest proportion of women with no education. 

Punjab has suffered from an imbalanced child sex ratio, starting already in the 1980's (908 girls per 1000 boys in 1981) when the child sex ratios were still normal in most other states in India. Rajasthan has remained as a state with a relatively high total fertility unlike the other states examined (TFR 4.1 in 1998). Kerala has enjoyed replacement level fertility since the early 1990's. Maharashtra has come to suffer from imbalance in child sex ratio during the last two decades, combined with replacement level fertility since the 2000's.

\subsection*{Appendix B: Likelihood conditioning on the sampling pattern}
To see that the likelihood obtained by multiplying (\ref{L2}) and (\ref{L3}) is still a conditional probability (less multiplicative terms), and thus a conditional likelihood, 
we partition the data collected under survey $j$ as $(v_j, w_j) \equiv \{(v_{ij}, w_{ij}): i \in C_j\}, \; j = 2, 3$, 
where $(v_{ij})$ represents the conditioning event or sampling pattern. Further,
$(w_{ij})$ denote the retrospective marriage histories recorded through the survey. Let $\Theta = (\theta, \beta(t), \mu(t,a))$ denote the parameters of interest $\theta$ 
as well as birth and mortality rates $(\beta(t), \mu(t,a))$. 
The parametrised joint distribution of all observed data $p(v_j, w_j, j = 2, 3 \mid \theta)$ may now be decomposed as 
\begin{align*}
p(v_2, w_2, v_3, w_3 \mid \Theta) &= p(w_2, w_3 \mid v_2, v_3 ; \Theta)p(v_2, v_3 \mid \Theta) \\
&= 
\prod_{j=2}^3 p(w_j \mid v_j; \theta)p(v_j \mid \Theta)\\
&= \prod_{j=2}^3 \prod_{i \in C_j} p(w_{ij} \mid v_{ij}; \theta) p(v_{ij} \mid \Theta) \\
&  \stackrel{\theta}{\propto} \prod_{j=2}^3 L_j(\theta) \prod_{j=2}^3 p(v_j \mid \Theta),
\end{align*}
where conditioning on the sampling plan (ignoring $\prod_{j=2}^3 p(v_j \mid \Theta)$) may 
result in some loss of information on $\theta$, but results in valid inferences.

\begin{table}[!h]
\small\sf\centering
\caption{Observed proportions (in $\%$) of women by state: categorical variables used are birth cohort, urban/rural, caste, religion, and education. (Source: NFHS-2 and -3 data)}
\begin{tabular}{c|ccccc}
\hline
 & Kerala & Maharashtra & Punjab & Rajasthan & India \\
 \hline
N & 6450 & 14424 & 6477 & 10701 & 214638 \\
\hline
Birth cohort   &     &    &   &    & \\
1942-1962  & 22  &15  &19 &18  &16 \\
1962-1972      & 33  &28  &30 &30  &28 \\
1972-1982      & 28  &33  &30 &36  &33 \\
1982-1992      & 17  &24  &21 &16  &23 \\
\hline
Urban & 33  & 66  & 36  & 28  & 40 \\
\hline
Caste  &     &     &    &    &  \\
SC     & 10  & 15  &30  &18  &17 \\
ST     & 1   & 8   &0.1 &14  &13 \\
OBC    & 38  & 25  &12  &31  &30 \\
Other  & 51  & 52  &58  &38  &40 \\
\hline
Religion  &     &     &    &     & \\
Hindu 	  & 55  & 75  &41  &89   &75 \\
Muslim    & 30  & 13  &3   &10   &13 \\
Christian & 15  & 2   &1   &0.1  &7 \\
Sikh      & 0   & 0.3 &55  &0.5  &2 \\
Other     & 0.1 & 10  &0.4 &1    &3 \\
\hline
Education  &    &    &    &    & \\
None       & 16 & 34 & 35 & 73 & 48  \\
Primary    & 39 & 34 & 30 & 18 & 28  \\
Secondary  & 31 & 20 & 26 & 6  & 16  \\
Higher     & 14 & 12 & 10 & 4  & 8  \\
\hline
\end{tabular}
\label{table:descriptivewomen}
\end{table}


\begin{thebibliography}{}

\bibitem[Cook and Lawless, 2018]{cook:2018}
Cook, R.J. and Lawless J.F. (2018).
\newblock {\em Multistate Models for the Analysis of Life History Data.}
\newblock Monographs on Statistics and Applied Probability 158, CRC Press.

\bibitem[Kashyap et~al., 2015]{kashyap:2015}
Kashyap, R., Esteve, A., and Garcia-Roman, J. (2015).
\newblock {{P}otential ({M}is)match? {M}arriage {M}arkets {A}midst
  {S}ociodemographic {C}hange in {I}ndia, 2005-2050}.
\newblock {\em Demography}, 52(1):183--208.

\bibitem[Keiding, 1991]{keiding:1991}
Keiding, N. (1991).
\newblock Age-specific incidence and prevalence: a statistical perspective.
\newblock {\em Journal of the Royal Statistical Society, Series A},
  154:371--412.

\bibitem[Keiding, 2006]{keiding:2006}
Keiding, N. (2006).
\newblock {Event history analysis and the cross-section.}  
\newblock {\em Statistics in Medicine}, 25:2343--2364.  
  
\bibitem[Keiding et~al., 2012]{keiding:2012}
Keiding, N., Hansen, O. K.~H., S{\o}rensen, D.~N., and Slama, R. (2012).
\newblock The current duration approach to estimating time to pregnancy.
\newblock {\em Scandinavian Journal of Statistics}, 39:185--204.

\bibitem[Mensch et al., 2005]{mensch:2005}
Mensch, B. S., Singh, S., Casterline, J. B. (2005).
\newblock {\em Trends in the Timing of First Marriage Among Men and Women in the Developing World}. 
\newblock The Population Council, Inc.


\bibitem[Ning et al., 2017]{ning:2017}
Ning, J., Hong, C., Li, L., Xuelin Huang, X. and Yu Shen, Y. (2017).
\newblock Estimating treatment effects in observational studies with both prevalent and incident cohorts
\newblock{\em The Canadian Journal of Statistics} 45:202–219.

\bibitem[Wolfson et al., 2019]{wolfson:2019}
Wolfson, D.B., Best,A.F., Addona, V., Wolfson, J. and Gadalla, S.M. (2019).
\newblock Benefits of combining prevalent and incident cohorts: An application to myotonic dystrophy
\newblock{\em Statistical Methods in Medical Research} 28:3333-3345.


\bibitem[{R Core Team}, 2020]{rmanual}
{R Core Team} (2020).
\newblock {\em R: A Language and Environment for Statistical Computing}.
\newblock R Foundation for Statistical Computing, Vienna, Austria.

\bibitem[Saarela et~al., 2009]{saarela:2009}
Saarela, O., Kulathinal, S., and Karvanen, J. (2009).
\newblock {{J}oint analysis of prevalence and incidence data using conditional
  likelihood}.
\newblock {\em Biostatistics}, 10(3):575--587.


\end{thebibliography}
\end{document}